\documentclass{article}
\usepackage{wrapfig}

\PassOptionsToPackage{numbers, compress}{natbib}

\usepackage[preprint]{neurips_2023}

\usepackage[utf8]{inputenc} 
\usepackage[T1]{fontenc}    
\usepackage{hyperref}       
\usepackage{url}            
\usepackage{booktabs}       
\usepackage{amsfonts}       
\usepackage{nicefrac}       
\usepackage{microtype}      
\usepackage{xcolor}         
\usepackage{caption}
\usepackage{amsmath}
\usepackage{graphicx}
\usepackage{xr}

\title{Expressive dynamics models with nonlinear injective readouts enable reliable recovery of latent features from neural activity}

\author{%
  Christopher Versteeg$^{1}$ \quad Andrew R. Sedler$^{1,2}$ \quad
  \textbf{Jonathan D. McCart$^{1,2}$} \\ \textbf{Chethan Pandarinath$^{1,2}$}\\
  \\
  $^1$ Wallace H. Coulter Department of Biomedical Engineering\\
  Emory University and Georgia Institute of Technology\\
  Atlanta, GA, USA\\
  \\
  $^2$ Center for Machine Learning\\
  Georgia Institute of Technology\\
  Atlanta, GA, USA
}
\begin{document}

\maketitle

\begin{abstract}
The advent of large-scale neural recordings has enabled new approaches that aim to discover the computational mechanisms of neural circuits by understanding the rules that govern how their state evolves over time. While these \textit{neural dynamics} cannot be directly measured, they can typically be approximated by low-dimensional models in a latent space. How these models represent the mapping from latent space to neural space can affect the interpretability of the latent representation. We show that typical choices for this mapping (e.g., linear or MLP) often lack the property of injectivity, meaning that changes in latent state are not obligated to affect activity in the neural space. During training, non-injective readouts incentivize the invention of dynamics that misrepresent the underlying system and the computation it performs. Combining our injective Flow readout with prior work on interpretable latent dynamics models, we created the Ordinary Differential equations autoencoder with Injective Nonlinear readout (ODIN), which learns to capture latent dynamical systems that are nonlinearly embedded into observed neural activity via an approximately injective nonlinear mapping. We show that ODIN can recover nonlinearly embedded systems from simulated neural activity, even when the nature of the system and embedding are unknown. Additionally, we show that ODIN enables the unsupervised recovery of underlying dynamical features (e.g., fixed points) and embedding geometry. When applied to biological neural recordings, ODIN can reconstruct neural activity with comparable accuracy to previous state-of-the-art methods while using substantially fewer latent dimensions. Overall, ODIN's accuracy in recovering ground-truth latent features and ability to accurately reconstruct neural activity with low dimensionality make it a promising method for distilling interpretable dynamics that can help explain neural computation.

\end{abstract}

\vspace{-3mm}
\section{Introduction}
\vspace{-3mm}

 Recent evidence has shown that when artificial recurrent neural networks are trained to perform tasks, the rules that govern how the internal activity evolves over time (i.e., the network dynamics) can provide insight into how the network performs the underlying computation \cite{sussillo_opening_2013, mante_context-dependent_2013, remington_flexible_2018, maheswaranathan_reverse_2019}. Given the conceptual similarities between artificial neural networks and biological neural circuits, it may be possible to apply these same dynamical analyses to brain activity to gain insight into how neural circuits perform complex sensory, cognitive, and motor processes \cite{vyas_computation_2020, shenoy_cortical_2013, jazayeri_interpreting_2021}. However, unlike in artificial networks, we cannot easily interrogate the dynamics of biological neural circuits and must first estimate them from observed neural activity. 

Fortunately, advances in recording technology have dramatically increased the number of neurons that can be simultaneously recorded, providing ample data for novel population-level analyses of neural activity \cite{stevenson_how_2011,steinmetz_neuropixels_2021, demas_high-speed_2021}. In these datasets, the activity of hundreds or thousands of neurons can often be captured by relatively low-dimensional subspaces \cite{gao_simplicity_2015}, orders-of-magnitude smaller than the total number of neurons. Neural activity in these latent spaces seems to evolve according to consistent sets of rules (i.e., latent dynamics) \cite{duncker_dynamics_2021, shenoy_cortical_2013}. Assuming no external inputs, these rules can be expressed mathematically as:
\begin{align}
    \mathbf{z}_{t+1} &= \mathbf{z}_{t}+f(\mathbf{z}_t) \\
    \mathbf{y}_t &= \exp g(\mathbf{z}_t) \\
    \mathbf{x}_t &\sim \text{Poisson}(\mathbf{y}_t)
\end{align}
where $\mathbf{z}_{t}\in \mathbb{R}^{D}$ represents the latent state at time $t$, $f(\mathbf{\cdot}): \mathbb{R}^{D}\rightarrow \mathbb{R}^{D}$ is the vector field governing the dynamical system, $\mathbf{y}_t\in \mathbb{R}^{N}$ denotes the firing rates of the $N$ neurons, $g(\cdot): \mathbb{R}^{D}\rightarrow \mathbb{R}^N$ maps latent activity into log-firing rates, and $\mathbf{x}_{t}\in \mathbb{R}^{N}$ denotes the observed spike counts at time $t$, assuming the spiking activity follows a Poisson distribution with time-varying rates given at each moment $t$ by $\mathbf{y}_t$.

Unfortunately, any latent system can be equivalently described by many combinations of dynamics $f$ and embeddings $g$, which makes the search for a unique latent system futile. However, versions of a latent system's dynamics $f$ and embedding $g$ that are less complex and use fewer latent dimensions can be  easier to interpret than alternative representations that are more complex and/or higher-dimensional. Models of latent dynamics that can discover simple and low-dimensional representations will make it easier to link latent dynamics to neural computation.

A popular approach to estimate neural dynamics \cite{sussillo_lfads_2016, schimel_ilqr-vae_2021, sedler_expressive_2023} is to use neural population dynamics models (NPDMs), which model neural activity as a latent dynamical system embedded into neural activity. We refer to the components of an NPDM that learn the dynamics and embedding as the generator $\hat{f}$ and the readout $\hat{g}$, respectively. When modeling neural activity, the  generator and readout are jointly trained to infer firing rates $\mathbf{\hat{y}}$ that maximize the likelihood of the observed neural activity $\mathbf{x}$.

Using NPDMs to estimate underlying dynamics and embedding implicitly assumes that good reconstruction performance (i.e., $\hat{\mathbf{x}} \approx \mathbf{x}$) implies interpretable estimates of the underlying system  (i.e., $\hat{\mathbf{z}} \approx \mathbf{z}$, $\hat{f} \approx f$, $\hat{g} \approx g$). However, recent work has shown that when the state dimensionality of the generator $\hat{D}$ is larger than a system's latent dimensionality $D$, high reconstruction performance may actually correspond to estimates of the latent system that are overly complex or misleading and therefore harder to interpret \cite{sedler_expressive_2023}. At present, reconstruction performance is seemingly an unreliable indicator for the interpretability of the learned dynamics. 

This vulnerability to learning overly complex latent features might emerge from the fact that, without constraints on the readout $\hat{g}$, changes in the latent state are not obligated to have an effect on predicted neural activity. Thus, NPDMs can be rewarded for inventing latent activity that boosts reconstruction performance, even if that latent activity has no direct correspondence to neural activity. A potential solution is to make $\hat{g}$ injective, which obligates all latent activity to affect neural reconstruction. This would penalize any latent activity that is not reflected in the observed neural activity, thereby putting pressure on the generator $\hat{f}$ and readout $\hat{g}$ to learn a more interpretable (i.e., simpler and lower dimensional) representation of the underlying system.

In addition, most previously used readouts $\hat{g}$ were not expressive enough to model diverse mappings from latent space to neural space, assuming the embedding $g$ to be a relatively simple (often linear) transformation (though there are exceptions \cite{gao_linear_2016,wu_gaussian_2017,zhao_variational_2020}). Capturing nonlinear embeddings is important because neural activity often lives on a lower-dimensional manifold that is nonlinearly embedded into the higher-dimensional neural space \cite{jazayeri_interpreting_2021}. Therefore, assumptions of linearity are likely to prevent NPDMs from capturing dynamics in their simplest and lowest-dimensional form, making them less interpretable than the latent features learned by NPDMs that can approximate these nonlinearities. 

To address these challenges, we propose a novel architecture called the Ordinary Differential equation autoencoder with Injective Nonlinear readout (ODIN), which implements $\hat{f}$ using a Neural ODE (NODE \cite{chen_neural_2019}) and $\hat{g}$ using a network inspired by invertible ResNets \cite{dinh2014nice, kingma2018glow, ardizzone_analyzing_2019, chen_neural_2019, behrmann_invertible_2019}. ODIN approximates an injective nonlinear mapping between latent states and neural activity, obligating all latent state variance to appear in the predicted neural activity and penalizing the model for using excessively complex or high-dimensional dynamics to model the underlying system. On synthetic data, ODIN learns representations of the latent system that are more interpretable, with simpler and lower-dimensional latent activity and dynamical features (e.g., fixed points) than alternative readouts. ODIN's interpretability is also more robust to overestimates of latent dimensionality and can recover the nonlinear embedding of synthetic data that evolves on a simulated manifold. When applied to neural activity from a monkey performing a reaching task with obstacles, ODIN reconstructs neural activity comparably to state-of-the-art recurrent neural network (RNN)-based models while requiring far fewer latent state dimensions. In summary, ODIN estimates interpretable latent features from synthetic data and has high reconstruction performance on biological neural recordings, making it a promising tool for understanding how the brain performs computation.
\section{Related Work}
\vspace{-3mm}

Many previous models have attempted to understand neural activity through the lens of neural dynamics. Early efforts limited model complexity by constraining both $\hat{f}$ and $\hat{g}$ to be linear \cite{macke_empirical_2011, archer_black_2015, pfau_robust_2013}. While these models were relatively straightforward to analyze, they often failed to adequately explain neural activity patterns \cite{pei_neural_2022}.

Other approaches increased the expressiveness of the modeled dynamics $\hat{f}$. RNNs can learn to approximate complex nonlinear dynamics, and have been shown to substantially outperform linear dynamics models in reconstructing neural activity \cite{pei_neural_2022}. Unfortunately, RNNs implicitly couple the capacity of the model to the latent state dimensionality, meaning their ability to model complex dynamics relies on having a high-dimensional latent state. In contrast, NODEs can model arbitrarily complex dynamics of embedded dynamical systems at the dimensionality of the system \cite{chen_neural_2019, sedler_expressive_2023}. On synthetic data, NODEs have been shown to recover dynamics more accurately than RNN-based methods \cite{kim2021inferring, sedler_expressive_2023}. In contrast to our approach, previous NODE-based models used a linear readout $\hat{g}$ that lacks injectivity. This can make the accuracy of estimated latent activity vulnerable to overestimates of the latent dimensionality (i.e., when $\hat{D}>D$) and/or fail to capture potential nonlinearities in the embedding $g$.

Early efforts to allow greater flexibility in $\hat{g}$ preserved linearity in $\hat{f}$, using feed-forward neural networks to nonlinearly embed linear dynamical systems in high-dimensional neural firing rates \cite{gao_linear_2016}. More recently, models have used Gaussian processes to approximate nonlinear mappings from latent state to neural firing with tuning curves \cite{wu_gaussian_2017}. Other models have combined nonlinear dynamics models and nonlinear embeddings for applications in behavioral tracking \cite{johnson_composing_2017} and neural reconstruction \cite{zhao_variational_2020}. Additional approaches extend these methods to incorporate alternative noise models that may better reflect the underlying firing properties of neurons \cite{gao_linear_2016, stevenson_flexible_2016}. While nonlinear, the readouts of these models lacked injectivity in their mapping from latent activity to neural activity. 

Many alternative models seek to capture interpretable latent features of a system from observations. One popular approach uses a sparsity penalty on a high-dimensional basis set to derive a sparse symbolic estimate of the governing equations for the system \cite{brunton2016discovering}. However, it is unclear whether such sparse symbolic representation is necessarily a benefit when modeling dynamics in the brain. Another recent model uses contrastive loss and auxiliary behavioral variables to learn low-dimensional representations of latent activity \cite{schneider_learnable_2023}. This approach does not have an explicit dynamics model, however, so is not amenable to the dynamical analyses performed in this manuscript.

Normalizing flows -- a type of invertible neural network -- have recently become a staple for generative modeling and density estimation \cite{dinh2014nice, behrmann_invertible_2019}. Some latent variable models have used invertible networks to approximate the mapping from the latent space to neural activity \cite{zhou_learning_2020} or for generative models of visual cortex activity \cite{bashiri_flow-based_2021}. To allow this mapping to change dimensionality between the latent space and neural activity, some of these models used a zero-padding procedure similar to the padding used in this manuscript (see Section \ref{Flow}), which makes the transformation injective rather than invertible \cite{zhou_learning_2020, behrmann_invertible_2019}. However, these previous approaches did not have explicit dynamics models, making our study, to our knowledge, the first to test whether injective readouts can improve the interpretability of neural population dynamics models.
\newpage
\section{Methods\label{Methods}}
\vspace{-3mm}
\subsection{Synthetic Neural Data\label{SynthData}}
\vspace{-3mm}
To determine whether different models can distill an interpretable latent system from observed population activity, we first used reference datasets that were generated using simple ground-truth dynamics $f$ and embedding $g$. Our synthetic test cases emulate the empirical properties of neural systems, specifically low-dimensional latent dynamics observed through noisy spiking activity \cite{sussillo_lfads_2016,smith_reverse_2021, hurwitz_targeted_2021, jensen_scalable_2021}. We sampled latent trajectories from the Arneodo system ($f$, $D=3$) and nonlinearly embedded these trajectories into neural activity via an embedding $g$ . We consider models that can recover the dynamics $f$ and embedding $g$ used to generate these data as providing an interpretable description of the latent system and its relation to the neural activity. Additional detail on data generation, models, and metrics can be found in the Supplementary Material.

\begin{figure}
    \centering
    \includegraphics[width=3.4in]{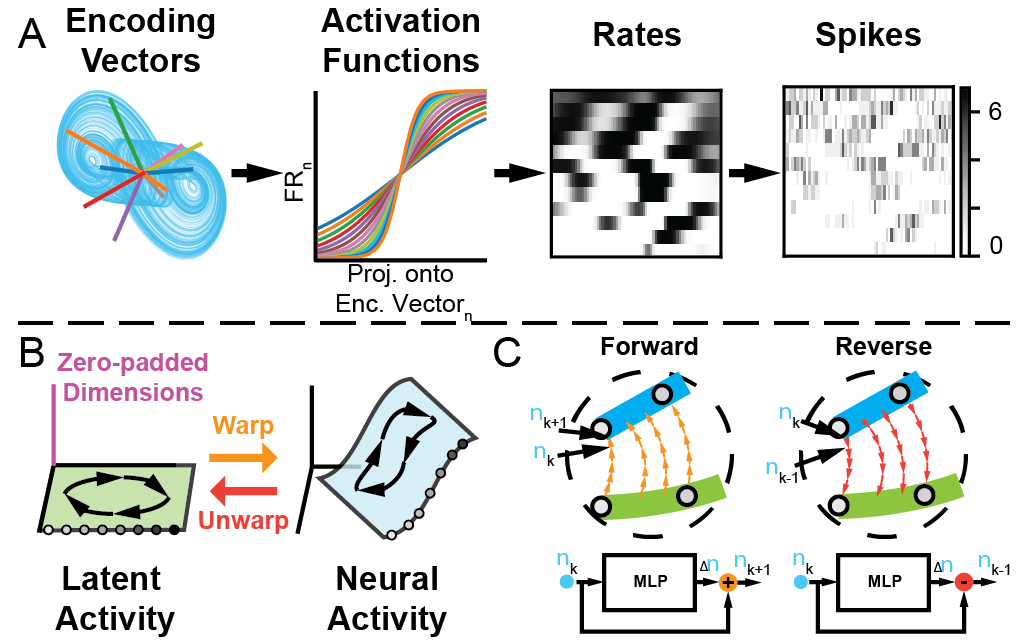}
    \caption{ A) Synthetic neural data generation (left to right). Trajectories from the Arneodo system are projected onto random encoding vectors to compute activations at each timepoint. A scaled sigmoid nonlinearity is applied to convert the activations into firing rates. B) Zero-padded latent dynamics (green) are reversibly warped into higher-dimensional neural activity space (blue). C) The Flow readout maps from latent space to neural space by applying a sequence of $K$ small updates (parameterized by an MLP, bottom). The reverse pass of the Flow maps from neural space to latent space and is implemented by serial subtraction of updates from the same MLP.}
    \label{fig:DataGenFlow}
    \vspace{-0.55cm}
\end{figure}

Unless otherwise noted, we generated activations for $N$ neurons ($N=12$) by projecting the simulated latent trajectories $\mathbf{Z}$ through a $3\times N$ matrix whose columns were random encoding vectors with elements sampled from a uniform distribution $U[-0.5, 0.5]$ (Fig. \ref{fig:DataGenFlow}A, left). We standardized these activations to have zero mean and unit variance and applied a different scaled sigmoid function to each neuron, yielding a matrix of non-negative time-varying firing rates $\mathbf{Y}$. The scaling of each sigmoid function was evenly spaced on a logarithmic scale between $10^{0.2}$ and $10$. This process created a diverse set of activation functions ranging from quasi-linear to nearly step-function-like behavior (Fig. \ref{fig:DataGenFlow}A, Activation Functions). For one experiment, we used the standard linear-exponential activation function, as described in previous work \cite{sedler_expressive_2023}, instead of the scaled sigmoid.

We simulated spiking activity $\mathbf{X}$ by sampling from inhomogeneous Poisson processes with time-varying rate parameters equal to the firing rates $\mathbf{Y}$ of the simulated neurons (Fig. \ref{fig:DataGenFlow}A, right). We randomly split 70-point segments of these trials into training and validation datasets (training and validation proportions were 0.8 and 0.2, respectively).
\vspace{-3mm}
\subsection{Biological Neural Data\label{BioData}}
\vspace{-3mm}
We evaluated how well our model could reconstruct biological neural activity on a well-characterized dataset \cite{churchland_cortical_2010} included in the Neural Latents Benchmark (NLB) \cite{pei_neural_2022}. This dataset is composed of single-unit recordings from primary and pre-motor cortices of a monkey performing a visually-guided reaching task with obstacles, referred to as the Maze task. Trials were trimmed to the window [-250, 350] ms relative to movement onset, and spiking activity was binned at 20 ms. To compare the reconstruction performance of our model directly against the benchmark, we split the neural activity into held-in and held-out neurons, comprising 137 and 35 neurons, respectively, using the same sets of neurons as were used to assess models for the NLB leaderboard.

\vspace{-3mm}
\subsection{Model Architecture\label{Arch}}
\vspace{-3mm}
We used three sequential autoencoder (SAE) variants in this study, with the main difference being the choice of readout module, $\hat{g}(\cdot)$. In brief, a sequence of binned spike counts $\mathbf{x}_{1:T}$ was passed through a bidirectional GRU encoder, whose final hidden states were converted to an initial condition $\hat{\mathbf{z}}_0$ via a mapping $\phi(\cdot)$. A modified NODE generator unrolled the initial condition into time-varying latent states $\hat{\mathbf{z}}_{1:T}$. These were subsequently mapped to inferred rates via the readout $\hat{g}(\cdot) \in \{\text{Linear}, \text{MLP}, \text{Flow}\}$. All models were trained for a fixed number of epochs to infer firing rates $\hat{\mathbf{y}}_{1:T}$ that minimize the negative Poisson log-likelihood of the observed spikes $\mathbf{x}_{1:T}$.
\begin{gather}
    \mathbf{h}_T = \left[\mathbf{h}_{fwd}\big| \mathbf{h}_{bwd}\right] = \text{BiGRU}(\mathbf{x}_{1:T}) \\
    \hat{\mathbf{z}}_0 = \phi(\mathbf{h}_T)  \\
    \hat{\mathbf{z}}_{t+1} = \hat{\mathbf{z}}_{t} + \alpha\cdot \text{MLP}(\hat{\mathbf{z}}_t)  \\
    \hat{\mathbf{y}}_t = \exp \hat{g}(\hat{\mathbf{z}}_t)
\end{gather}
For models with Linear and MLP readouts, $\phi(\cdot)$ was a linear map to $\mathbb{R}^{\hat{D}}$. For models with Flow readouts, $\phi(\cdot)$ was a linear map to $\mathbb{R}^N$ followed by the reverse pass of the Flow (see Section \ref{Flow}). We unrolled the NODE using Euler's method with a fixed step size equal to the bin width and trained using standard backpropagation for efficiency. A scaling factor ($\alpha=0.1)$ was applied to the output of the NODE's MLP to stabilize the dynamics during early training. Readouts were implemented as either a single linear layer (Linear), an MLP with two 150-unit ReLU hidden layers (MLP), or a Flow readout (Flow) which contains an MLP with two 150-unit ReLU hidden layers. We refer to these three models as Linear-NODE, MLP-NODE, and ODIN, respectively. 

\vspace{-3mm}
\subsubsection{Flow Readout\label{Flow}}
\vspace{-3mm}

The Flow readout resembles a simplified invertible ResNet \cite{behrmann_invertible_2019}. Flow learns a vector field that can reversibly transform data between latent and neural representations (Figure \ref{fig:DataGenFlow}B). The Flow readout has three steps: first, we increase the dimensionality of the latent activity $\mathbf{z}_t$ to match that of the neural activity by padding the latent state with zeros. This corresponds to an initial estimate of the log-firing rates, $\log \hat{\mathbf{y}}_{t,0}$. Note that zero-padding makes our mapping injective rather than fully invertible (see \cite{ behrmann_invertible_2019, zhou_learning_2020}). The Flow network then uses an MLP to iteratively refine $\log \hat{\mathbf{y}}_{t,k}$ over $K$ steps ($K=20$) after which we apply an exponential to produce the final firing rate predictions, $\hat{\mathbf{y}}_t$. A scaling factor ($\beta=0.1)$ was applied to the output of the Flow's MLP, which prevents the embedding from becoming unstable during the early training period.
\begin{gather}
    \log \hat{\mathbf{y}}_{t,0} = [\hat{\mathbf{z}}_t | \mathbf{0}]^T \\
    \log \hat{\mathbf{y}}_{t,k+1} = \log \hat{\mathbf{y}}_{t,k} + \beta \cdot \text{MLP}(\log \hat{\mathbf{y}}_{t,k}) \\
    \hat{g}\left(\hat{\mathbf{z}}_t\right) = \log \hat{\mathbf{y}}_{t,K} = \log \hat{\mathbf{y}}_t
\end{gather}
We also use a reverse pass of the Flow to transform the output of the encoders to initial conditions in the latent space via $\phi(\cdot)$, approximating the inverse function $\hat{g}^{-1}$. Our method subtracts the output of the MLP from the state rather than adding it as in the forward mode (Fig \ref{fig:DataGenFlow}C), a simplified version of the fixed-point iteration procedure described in \cite{behrmann_invertible_2019}. We then trim the excess dimensions to recover $\hat{z}\in \mathbb{R}^{\hat{D}}$ (in effect, removing the zero-padding dimensions).
\begin{gather}
    \log \hat{\mathbf{y}}_{t,k-1} = \log \hat{\mathbf{y}}_{t,k} - \beta \cdot \text{MLP}(\log \hat{\mathbf{y}}_{t,k}) \\
    \hat{g}^{-1}\left(\log \hat{\mathbf{y}}_{t}\right) = [\log \hat{y}_{t,0,1}, \dots, \log \hat{y}_{t,0,\hat{D}}]^T  = \hat{\mathbf{z}}_t
\end{gather}
The Flow mapping is only guaranteed to be injective if changes in the output of the MLP are sufficiently small relative to changes in the input (i.e., Lipschitz constant for the MLP that is strictly less than 1) \cite{behrmann_invertible_2019}. The model can be made fully injective by either restricting the weights of the MLP (e.g., spectral norm \cite{miyato_spectral_2018}), or using a variable step-size ODE solver that can prevent crossing trajectories (e.g., continuous normalizing flows \cite{chen_neural_2019}). In practice, we found that using a moderate number of steps allows Flow to preserve approximate injectivity of the readout at all tested dimensionalities (Supp. Fig. S2).

\vspace{-3mm}
\subsection{Metrics and characterization of dynamics\label{Metrics}}
\vspace{-3mm}
\captionsetup[figure]{skip=-5pt} 

We assessed model performance in five domains: 1) reconstruction performance, 2) latent accuracy, 3) dynamical accuracy, 4) embedding accuracy, and 5) readout injectivity. All metrics were evaluated on validation data. Critically, on biological data without a ground-truth system, only the reconstruction performance and readout injectivity can be assessed, since all the other metrics rely on full observability of the underlying system. Therefore, we need models for which good performance on the observable metrics (reconstruction, injectivity) implies good performance on the unobservable metrics (latent, dynamical, and embedding accuracy).

Reconstruction performance for the synthetic data was assessed using two key metrics. The first, spike negative log-likelihood (Spike NLL), was defined as the Poisson NLL employed during model training. The second, Rate $R^2$, was the coefficient of determination between the inferred and true firing rates, averaged across neurons. We used Spike NLL to assess how well the inferred rates explain the spiking activity, while Rate $R^2$ reflects the model's ability to find the true firing rates. These metrics quantify how well the model captures the embedded system's dynamics (i.e., that $\hat{f}, \hat{g}$ captures the system described by $f,g$), but give no indication of the interpretability of the learned latent representation (i.e., that the learned $\hat{f}, \hat{g}$ are simple and low-dimensional).

For the biological neural data, we measured model performance using two metrics from the Neural Latents Benchmark (NLB) \cite{pei_neural_2022}, co-smoothing bits-per-spike (co-bps) and velocity decoding performance on predicted firing rates (Vel $R^{2}$). co-bps is a measure of reconstruction performance that quantifies how well the model predicts the spiking of the held-out neurons, while Vel $R^{2}$ quantifies how well the denoised rates can predict the monkey's hand velocity during the reach. We have no way to directly assess embedding, latent, or dynamical accuracy because they are unobserved in most biological datasets.

To determine whether a model's inferred latent activity contains features that are not in the simulated latent activity, we used a previously published metric called the State $R^2$ \cite{sedler_expressive_2023}. State $R^2$ is defined as the coefficient of determination ($R^2$) of a linear regression from simulated latent trajectories $\mathbf{z}$ to the inferred latent trajectories $\hat{\mathbf{z}}$. State $R^2$ will be low if the inferred latent trajectories contain features that cannot be explained by an affine transformation of the true latent trajectories. Importantly, State $R^2$ alone cannot ensure latent accuracy. This is because a model can achieve high State $R^2$ trivially if the inferred latent activity $\hat{\mathbf{z}}$ is a low-dimensional projection of the simulated activity $\mathbf{z}$. Therefore, only models that have \emph{both} good reconstruction performance (Spike NLL, Rate $R^2$) and State $R^2$ can be said to accurately reflect the simulated latent dynamics without extra features that make the model harder to interpret (i.e., $\hat{\mathbf{z}} \approx \mathbf{z}$).

As a direct comparison of the estimated dynamics $\hat{f}$ to the simulated dynamics $f$, we extracted the fixed-point (FP) structure from our trained models and compared it to the FP structure of the underlying system. We used previously published FP-finding techniques \cite{Golub2018} to identify regions of the generator's dynamics where the magnitude of the vector field was close to zero, calling this set of locations the putative FPs. We linearized the dynamics around the FPs and computed the eigenvalues of the Jacobian of $\hat{f}$ to characterize each FP. Capturing FP location and character gives an indication of how closely the estimated dynamics resemble the simulated dynamics (i.e., $\hat{f} \approx f$).

To determine how well our embedding $\hat{g}$ captures the simulated embedding $g$, we projected the encoding vectors used to generate the synthetic neural activity from the ground-truth system into our model's latent space using the same affine transformation from ground-truth latent activity to inferred latent activity that was used to compute State $R^2$. We projected the inferred latent activity onto each neuron's affine-transformed encoding vector to find the predicted activation of each synthetic neuron. We then related the predicted firing rates of each neuron to its corresponding activations to derive an estimate of each neuron's activation function. Because the inferred latent activity is arbitrarily scaled/translated relative to the true latent activity, we fit an affine transformation from the predicted activation function to the ground-truth activation function. The coefficient of determination $R^2$ of this fit quantifies how well our models were able to recover the synthetic warping applied to each neuron (i.e., $\hat{g} \approx g$).

We compared the injectivity of the Flow readout to Linear and MLP readouts using effective rank \cite{roy_effective_2007} and cycle-consistency, respectively. Effective rank quantifies the number of significant singular values in a Linear readout, while cycle-consistency quantifies how well the inferred latent activity $\hat{\mathbf{z}}$ can be recovered from the predicted log-firing rates $\log \hat{\mathbf{y}}$.

\begin{figure*}[h]
    \centering
    \includegraphics[width = \textwidth]{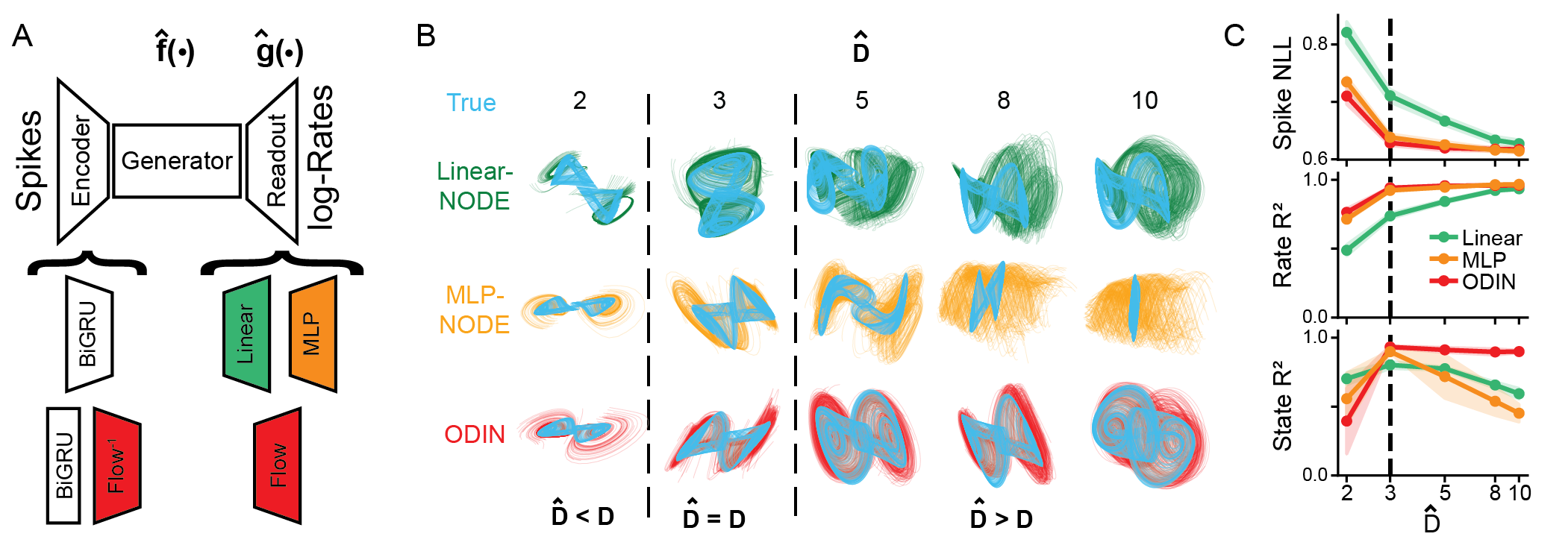}
    \caption{ ODIN recovers latent activity more accurately than alternative models and is robust to overestimates of latent dimensionality. A) Diagram of models tested, including Linear-NODE (green), MLP-NODE (orange), ODIN (red). B) Inferred latent activity of representative model at each state dimensionality $\hat{D}$. True latent activity (affine-transformed to overlay inferred latent activity) shown in light blue. C) All: Model metrics as a function of $\hat{D}$. Shaded areas represent one standard deviation around the mean. Dashed vertical line indicates $\hat{D}=3$ Top: Spike NLL, Middle: Rate $R^2$, Bottom: State $R^2$.}
    \label{fig:flow-mlp-lin}
\end{figure*}
\newpage
\vspace{-3mm}
\section{Results\label{Results}}
\vspace{-3mm}

\subsection{Finding interpretable latent activity across state dimensionalities with ODIN\label{LatAccuracySection}}
\vspace{-3mm}
As the latent dimensionality $D$ is unknown for biological datasets, we wanted to test how robust each model was to choices of state dimensionality $\hat{D}$. We trained Linear/MLP -NODE, and ODIN (Fig \ref{fig:flow-mlp-lin}A) to reconstruct synthetic neural activity from the Arneodo system \cite{arneodo_occurence_1980} and compared reconstruction performance (i.e. Spike NLL and Rate $R^2$) and latent recovery (i.e. State $R^2$) as functions of the dimensionality $\hat{D}$ of the state space. We trained 5 different random seeds for each of the 3 model types and 5 state dimensionalities (75 total models, model hyperparameters in Supp. Table 1, representative hyperparameter sweeps in Supp. Fig. S1). 

First, we observed that latent activity inferred by Linear-NODE did not closely resemble the simulated latent activity, with all tested dimensionalities performing worse than either ODIN or the MLP-NODE at $\hat{D}$ = 3 (Fig \ref{fig:flow-mlp-lin}B,C, mean State $R^2$ = 0.70 for Linear-NODE vs. 0.89, 0.93 for MLP-NODE, ODIN respectively). We also found that Linear-NODE required many more dimensions to reach the peak reconstruction performance (Fig \ref{fig:flow-mlp-lin}C, Rate $R^2$).  These results demonstrate that models that are unable to account for nonlinear embeddings are vulnerable to learning more complex and higher dimensional dynamics than those learned by models with nonlinear readouts.

Next, we compared ODIN to MLP-NODE and found that at the correct dimensionality ($\hat{D} = 3$), these models had similar performance for both reconstruction and latent recovery. However, as the dimensionality increased beyond the true dimensionality ($\hat{D} > 3$), the latent recovery of the MLP-NODE degraded rapidly while ODIN's latent recovery remained high (Fig \ref{fig:flow-mlp-lin}C, as $\hat{D}>3$). As the true latent dimensionality $D$ is usually unknown, NPDMs with non-injective readouts (like MLPs) may be predisposed to learning misleading latent activity that can make it more difficult to interpret biological datasets. 

\vspace{-3mm}
\subsection{Common readouts learn non-injective mappings from latent activity to firing rates \label{InjDemo}}
\vspace{-3mm}

We then sought to assess the injectivity of different readouts. First, we used effective rank \cite{roy_effective_2007} to quantify the injectivity of our Linear readouts. We trained 5 Linear-NODE models at a range of state dimensionalities ($\hat{D}=3,5,8,10$) to reconstruct simulated neural activity from Arneodo that was \textit{linearly} embedded into 12D neural space. We found that while reconstruction performance was optimal when $\hat{D}>3$ (Supp. Fig. S3), the effective rank of these best-reconstructing models never exceeded 4 (mean erank = 3.74 at $\hat{D}=10$). This means that for the largest Linear-NODE models, around 6 of 10 latent dimensions had no effect on reconstructed log-rates. The fact that linear readouts learn mappings with low effective rank, coupled with improved reconstruction performance when $\hat{D}>3$ suggests that the Linear readouts utilize non-injectivity to improve reconstruction at the expense of latent accuracy.
\newpage
\begin{wrapfigure}[27]{r}{2in}
    \begin{center}
    \includegraphics[width=1.5in, height=2.5in]{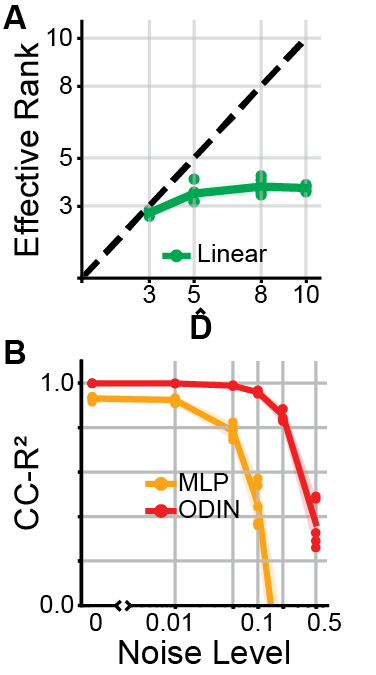}
    \end{center}
    \caption{Linear- and MLP-NODEs tend towards non-injectivity A) Effective rank of Linear readout as a function of state dimensionality $\hat{D}$. Each point represents one randomly instantiated model. B) Cycle-consistency $R^2$ for ODIN and MLP-NODE as a function of noise corruption.}
    \label{fig:Injectivity}
    \vspace{-0.2cm}
\end{wrapfigure}
Next, we used a cycle consistency metric to show that MLP readouts also have a tendency to become non-injective. Cycle consistency quantifies how well inputs to a function can be recovered from the function's outputs. We trained a separate MLP to predict inferred latents $\hat{\mathbf{z}}$ from predicted log-firing rates $\log \hat{\mathbf{y}}$ for 10D MLP-NODE and ODIN models shown in Figure \ref{fig:flow-mlp-lin}. We found that the cycle consistency of the ODIN model was consistently higher than for MLP-NODE (Fig. \ref{fig:Injectivity}B, Noise Level = 0). It is possible that models may learn to compress latent activity to arbitrarily small firing rate changes while still remaining technically injective. This failure mode could potentially be invisible to the standard cycle-consistency. To address this concern, we added Gaussian noise to the log-firing rates $\log \hat{\mathbf{y}}$ and tried to recover the inferred latent activity from these noise corrupted log-rates. Consistent with ODIN's bias towards injectivity, we found that ODIN's cycle consistency was more robust to the addition of noise than MLP-NODE (Fig. \ref{fig:Injectivity}B, Noise Level > 0).

To demonstrate that injectivity was the critical feature that allowed ODIN to outperform other models, we tested an alternative injective readout, an Invertible Neural Network (INN). INN implementation differs significantly from Flow, but they share the property of injectivity. We found that INN-NODE qualitatively reproduced ODIN's performance in Figure \ref{fig:flow-mlp-lin}C (Supp. Fig. S4), suggesting that the injectivity is the critical feature for recovering interpretable latent activity. We describe the advantages of ODIN over INN-NODE in the Supplementary material.

\vspace{-3mm}
\subsection{Recovering fixed point structure with ODIN\label{FPRecoverySection}}
\vspace{-3mm}

\begin{figure}[h!]
    \centering
    \includegraphics[width=3.46in]{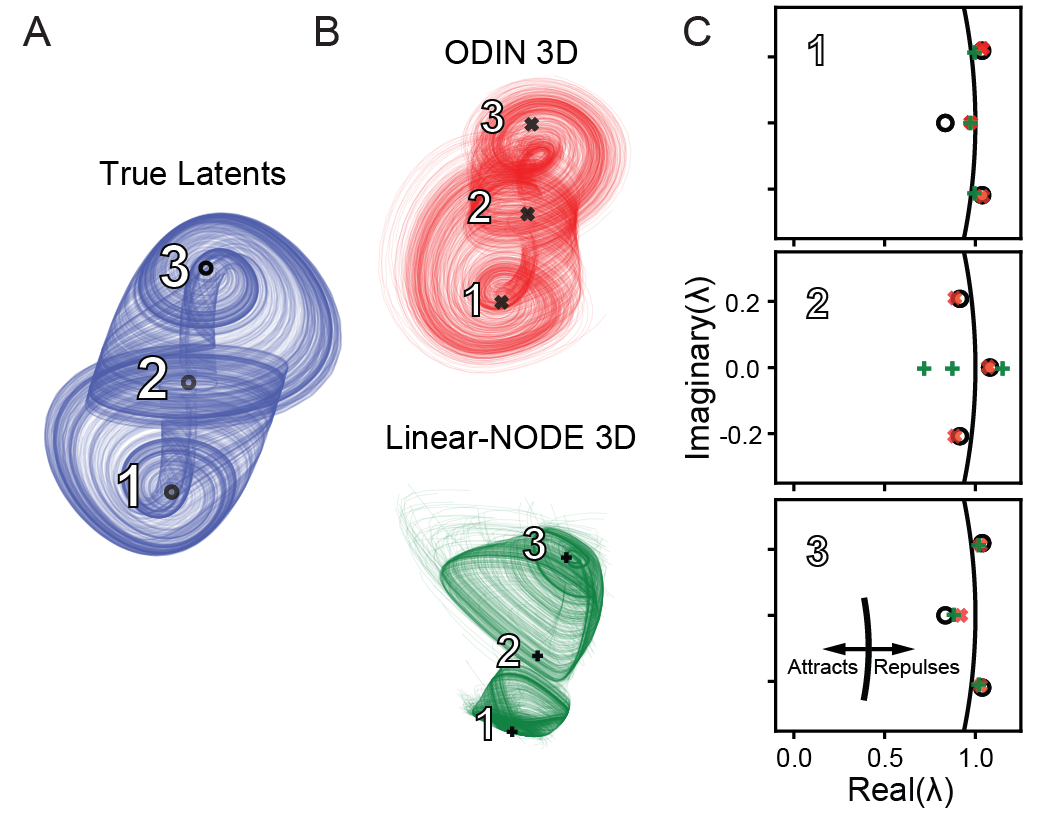}
    \caption{ ODIN recovers fixed point properties accurately at the correct dimensionality. A,B) Representative latent activity and fixed-points from the true (blue, $\circ$), ODIN (red, $\times$), and Linear-NODE (green, $+$) systems. Each fixed point is labeled with reference to C. C) Plots of the real vs. imaginary part of the eigenvalues of the Jacobian evaluated at each fixed point. Unit circle in the complex plane (black curve) shows boundary between attractive and repulsive behavior (the attractive and repulsive sides of the boundary are indicated by inset).}
    \label{fig:FP recovery}
    \vspace{-0.25cm}
\end{figure}

A common method to examine how well dynamics models capture the underlying dynamics from synthetic data is to compare the character and structure of the inferred fixed points (FPs) to the FPs of the ground-truth system \cite{sedler_expressive_2023}. At a high-level, FPs enable a concise description of the dynamics in a small region of state-space around the FP, and can collectively provide a qualitative picture of the overall dynamical landscape. To obtain a set of candidate FPs, we searched the latent space for points at which the magnitude of the vector field $\lVert \hat{f} \rVert$ is minimized (as in \cite{sussillo_opening_2013, Golub2018}). We computed the eigenvalues of the Jacobian of $\hat{f}$ at each FP location. The real and imaginary components of these eigenvalues identify each FP as attractive, repulsive, etc. 

We found that 3D ODIN models and 3D Linear-NODEs were both able to recover three fixed points that generally matched the location of the three fixed points of the Arneodo system (Fig \ref{fig:FP recovery}A), However, while ODIN was also able to capture the eigenspectra of all three FPs (Fig. \ref{fig:FP recovery}B, red $\times$), the Linear-NODE failed to capture the rotational dynamics of the central FP (Fig \ref{fig:FP recovery}B, middle column, green $+$). Both models were able to approximately recover the eigenspectra of outermost FPs of the system (Fig. \ref{fig:FP recovery}B, left, right columns). We found that the MLP-NODE was also able to find FPs with similar accuracy to ODIN at 3D. These results show that the inability to model the nonlinear embedding can lead to impoverished estimates of the underlying dynamics $\hat{f}$.

\vspace{-3mm}
\subsection{Recovering simulated activation functions with ODIN\label{ActRecoverySection}}
\begin{wrapfigure}[36]{r}{2.1in}
    \begin{center}
    \includegraphics[width=2.1in, height=3.1in, trim =10 0 10 20]{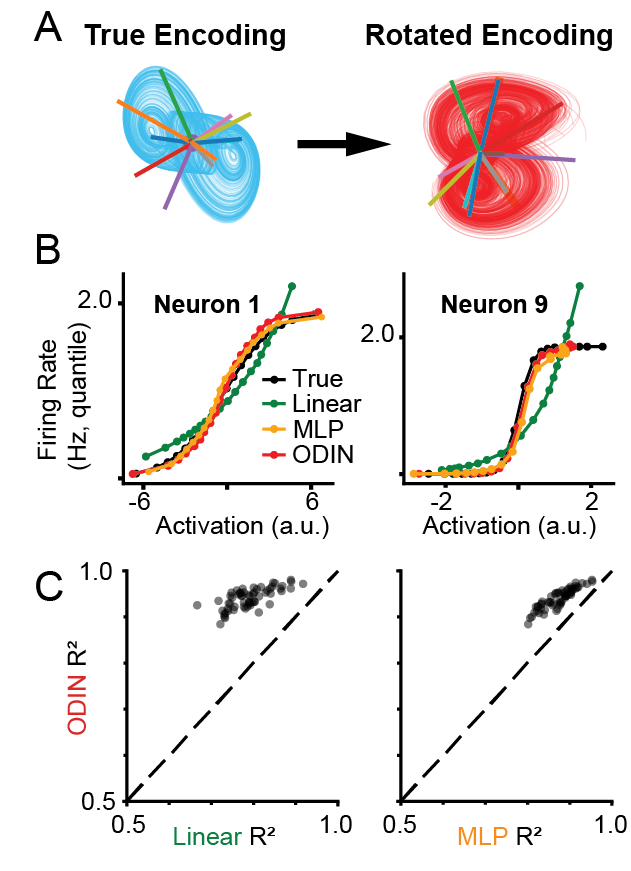}
    \end{center}
    \caption{ODIN can recover nonlinear activation functions of neurons. A) True encoding vectors (numbered lines over true latent activity (blue)) were affine-transformed into a representative model's latent space. B) Inferred activation function for two example neurons (columns), color coded by readout type (Linear-NODE = green, MLP-NODE = orange, ODIN  = red, True = black). Plots show the predicted firing rate vs. the activation of the selected neuron. C) Comparison of the $R^2$ values of the fits across all neurons for models with $\hat{D} =3$.}
    \label{fig:ActRecovery}
\end{wrapfigure}
While obtaining interpretable dynamics is our primary goal, models that allow unsupervised recovery of the embedding geometry may provide additional insight about the computations performed by the neural system \cite{gardner_toroidal_2021, jazayeri_interpreting_2021}. For this section, we considered a representative model from each readout class with the correct number of latent dimensions ($D=3$). We performed an affine transformation from the ground truth encoding vectors into the modeled latent space and computed the projection of the modeled latent activity onto the affine-transformed encoding vectors (Fig \ref{fig:ActRecovery}A). From this projection, we derived an estimate of the activation function for each neuron, and compared this estimate to the ground-truth activation function. 

We found, as expected, that Linear-NODE was unable to approximate the sigmoidal activation function of individual neurons (Fig \ref{fig:ActRecovery}B, green). On the other hand, both ODIN and MLP-NODE were able to capture activation functions ranging from nearly linear to step function-like in nature (Fig \ref{fig:ActRecovery}B, red, orange). Across all simulated neurons for models with $D=3$, we found that ODIN more accurately estimated the activation function of individual neurons compared to both Linear- and MLP-NODEs (Fig \ref{fig:ActRecovery}C), suggesting that ODIN's injectivity allows more accurate estimation of nonlinear embeddings (two-sided paired t-test, p-val for ODIN vs. Linear-, MLP-NODE < 1e-10).

\vspace{-3mm}
\subsection{Modeling motor cortical activity with ODIN \label{MazeSection}}
\vspace{-3mm}

\begin{figure*}[h]
    \centering
    \includegraphics[width = 5.5in]{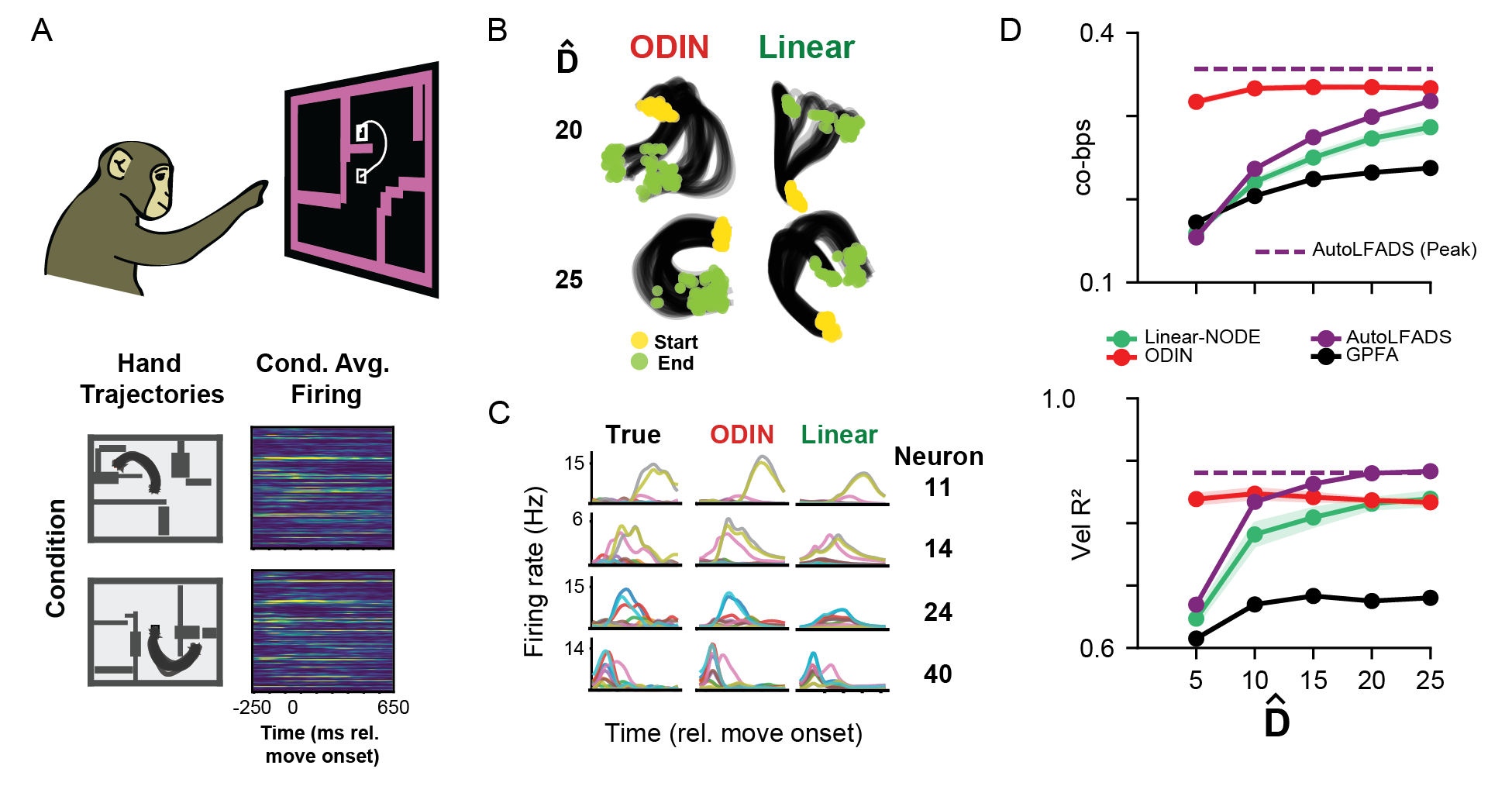}
    \caption{ ODIN can reconstruct cortical activity with low-dimensional dynamics A) Top: Schematic of task \cite{churchland_cortical_2010} Bottom: example hand trajectories and condition-averaged firing rates aligned to move onset. B) Example condition-averaged latent activity from ODIN and Linear-NODE models applied to neural activity recorded during the Maze task. C) Example single-neuron peri-stimulus time histograms for ODIN and Linear-NODE models across conditions. D) Effects of latent state dimensionality $\hat{D}$ on reconstruction (top, co-bps) and decoding (bottom, Vel $R^2$) performance. Plot shows mean (point) and standard deviation (shading) of 5 randomly initialized ODIN and Linear-NODE models at each $\hat{D}$. GPFA and AutoLFADS were a single run, or the best performing model from an adaptive hyperparameter search, respectively. Horizontal lines represent peak performance by AutoLFADS with $\hat{D}=100$.}
    \label{fig:Maze}
    \vspace{-0.25cm}
\end{figure*}

To validate ODIN's ability to fit neural activity from a biological neural circuit, we applied ODIN to the Maze dataset from the Neural Latents Benchmark, composed of recordings from the motor and pre-motor cortices of a monkey performing a reaching task (Fig. \ref{fig:Maze}A). After performing hyperparameter sweeps across regularization parameters and network size (Supp. Table 2), we trained a set of ODIN and Linear-NODE models to reconstruct the neural activity with a range of state dimensionalities $\hat{D}$. We visualized the top 3 PCs of the condition-averaged latent trajectories and predicted single-neuron firing rates for example models from each readout type. We found no visually obvious differences in the inferred latent trajectories (Fig. \ref{fig:Maze}B), but when we computed condition-averaged peri-stimulus time histograms (PSTHs) of single neuron firing rates, we found that ODIN typically produced firing rate estimates that more closely resembled the empirical PSTHs than those from the Linear-NODE (Fig. \ref{fig:Maze}C). 

Without access to a ground truth dynamics $f$ and embedding $g$ that generated these biological data, the dimensionality required to reconstruct the neural activity was our primary measure of interpretability. We computed co-bps --a measure of reconstruction performance on held-out neurons-- for each model and found that 10D ODIN models substantially outperformed Linear-NODE models, even when the Linear-NODE had more than twice as many dimensions (10D ODIN: 0.333, vs 25D Linear: 0.287). This suggests that ODIN's injective non-linear readout is effective at reducing the state dimensionality required to capture the data relative to a simple linear readout. 

We also compared ODIN to alternative models including AutoLFADS, GPFA, and MLP-NODE \cite{pei_neural_2022} at the same state dimensionalities. Trained AutoLFADS and GPFA models had lower co-bps at all tested state dimensionalities. In particular, co-bps was substantially higher for 10D ODIN compared to the 10D AutoLFADS or GPFA models (0.333 vs. 0.237, 0.204, respectively). As expected, MLP-NODE (not shown) performed similarly to ODIN; however, without a known state dimensionality, the MLP readout may incentivize the MLP-NODE to invent latent activity that is not reflected in the dataset. Of note, increasing AutoLFADS to a very high state dimensionality ($\hat{D}=100)$ allowed it to outperform ODIN in co-bps. However, as we have shown in Figures  \ref{fig:flow-mlp-lin} and \ref{fig:Injectivity}, improved reconstruction performance often comes at the expense of accuracy in latent recovery. Together, these results suggest that ODIN is effective at reducing the state dimensionality needed for good neural reconstruction, which may provide more interpretable latent representations than alternative models.
\vspace{-3mm}

\vspace{-3mm}
\section{Discussion\label{Discussion}}
\vspace{-3mm}

Dynamics models have had great success in reproducing neural activity patterns and relating brain activity to behavior \cite{pandarinath_inferring_2018, pei_neural_2022, smith_simplified_2023}. However, it has been difficult to use these models to investigate neural computation directly. If neural population models could be trusted to find interpretable representations of latent dynamics, then recent techniques that can uncover computation in artificial networks could help to explain computations in the brain \cite{sussillo_opening_2013, Golub2018, driscoll_flexible_2022}. In this work, we created a new model called ODIN that can overcome major barriers to learning interpretable latent dynamical systems. By combining Neural ODE generators and approximately injective nonlinear readouts, ODIN offers significant advantages over the current state-of-the-art, including lower latent dimensionality, simpler latent activity that is robust to the choice of latent dimensionality, and the ability to model arbitrary nonlinear activation functions.

Circuits in the brain are densely interconnected, and so a primary limitation of this work is that ODIN is not yet able to account for inputs to the system that may be coming from areas that are not directly modeled. Thus ODIN is currently only able to model the dynamics of a given population of neurons as an autonomous system. Inferring inputs is difficult due to ambiguity in the role and timecourse of inputs compared to internal dynamics for driving the state of the system. While some RNN-based models have methods for input inference \cite{pandarinath_inferring_2018}, more work is needed to develop solutions for NODE-based models. Injective readouts are an important step towards addressing the fundamental difficulties of input inference, as models without injective readouts can be incentivized to imagine latent features that are actually the result of inputs.

Interpretable dynamics derived from neural population recordings could answer critical scientific questions about the brain and help improve brain-machine interface technology. A potential negative consequence is that human neural interfaces combined with an understanding of neural computation might make it possible and profitable to develop strategies that are effective at influencing behavior. Future researchers should focus on applications of this research that are scientific and medical rather than commercial or political. 
\vspace{-3mm}
\section{Acknowledgements\label{Acks}}
\vspace{-3mm}

The authors would like to acknowledge Timothy D. Kim and Carlos Brody for helpful discussions that further developed the ideas in this manuscript.

This work was supported by NSF NCS 1835364, NIH-NINDS/OD DP2NS127291, NIH BRAIN/NIDA RF1 DA055667, and the Alfred P. Sloan Foundation (CP), NIH BRAIN/NINDS F32 RFA-MH-23-110 (CV), the Simons Foundation as part of the Simons-Emory International Consortium on Motor Control (CP, CV), and NSF Graduate Research Fellowship DGE-2039655 (ARS).

\newpage
\bibliographystyle{unsrtnat}
\bibliography{refs}

\medskip

\newpage

\setlength{\parindent}{0em}
\setlength{\parskip}{1em}
\renewcommand\thesection{\Alph{section}}

\emergencystretch 3em
\newcommand{\beginsupplement}{
    \setcounter{table}{0}
    \renewcommand{\thetable}{S\arabic{table}}
    \setcounter{figure}{0}
    \renewcommand{\thefigure}{S\arabic{figure}}
}
\beginsupplement

\newcommand{\answerYes}[1][]{\textcolor{blue}{[Yes] #1}}
\newcommand{\answerNo}[1][]{\textcolor{orange}{[No] #1}}
\newcommand{\answerNA}[1][]{\textcolor{gray}{[N/A] #1}}
\newcommand{\answerTODO}[1][]{\textcolor{red}{\bf [TODO]}}

\begin{center}
  \Large\textbf{Expressive dynamics models with nonlinear injective readouts enable reliable recovery of latent features from neural activity} \\
  \Large\textit{Supplementary Material}
\end{center}

\section{Datasets}

\subsection{Simulated neural data \label{synthetic_neural_data-Arneodo}}

\subsubsection{Latent trajectories\label{Arneodo-trajectories}}

We used the Arneodo system \cite{arneodo_occurence_1980} to generate synthetic data because it exhibits mildly chaotic behavior (Lyapunov exponent equal to 0.243), it has a low-dimensional state space, and the regions around its fixed points are well-sampled by trajectories of the system. As demonstrated by \cite{sedler_expressive_2023}, these properties allow recovery of latent dynamics in the absence of a nonlinear embedding. The Arneodo system is described by the following system of equations
\begin{align}
    \dot{x} &= y \\
    \dot{y} &= z \\
    \dot{z} &= -a x - b y - c z +d x^3
\end{align}

where $a=-5.5$, $b=4.5$, $c=1.0$, and $d=-1.0$ \cite{arneodo_occurence_1980}.

The system was simulated using the \texttt{dysts} Python package, which offered well-reasoned standards for initial conditions, integration steps, and resampling frequency \cite{gilpin_chaos_2021}. Initial conditions had been determined by running the model until the moments of the autocorrelation function were stationary. Integration steps had been chosen based on the highest significant frequency observed in the power spectrum. After integration, trajectories were resampled to contain 35 points per period, where period was based on the dominant frequency in the power spectrum. 

\subsubsection{Embedding low-dimensional trajectories on a nonlinear manifold}
We simulated neural activity by nonlinearly embedding the Arneodo trajectories as firing rates in the neural space. First, the trajectories were linearly projected into the neural space via a set of encoding vectors $\mathbf{\gamma}_i$ and standardized for each neuron (see Methods). These activations $\mathbf{a}_{i}$ were passed through a sigmoid with input scaling $\mathbf{\eta}_i$ and output scaling $b=2$ to produce reasonable firing rates as follows:
\begin{gather}
    \mathbf{\eta_i} = 10^{0.8\cdot\frac{i-1}{N-1} + 0.2},\\
    \mathbf{y}_{i} = \psi_i(\mathbf{a}_{i}) = b \times \sigma(\mathbf{\eta}_i \times \mathbf{a}_{i}), \quad i = 1, 2, \dots, N.
    \label{act_func}
\end{gather}
where $\sigma (\cdot)$ denotes the sigmoid function. This resulted in a set of activation functions $\psi _i (\cdot)$ ranging from quasi-linear to step-like. The resulting rates $\mathbf{y}_{i}$ were used to parameterize a Poisson process, which was sampled to obtain spiking data for $N$ neurons ($N=12$).

\subsubsection{Embedding low-dimensional trajectories onto linear manifold}
For Figure 2A, we tested whether Linear-NODEs fit to linearly-embedded data would find non-injective readouts when $\hat{D}>D$. We simulated an alternative dataset with the same procedure as above, except instead of passing the activations $\mathbf{a}_i$ through the sigmoidal non-linearity, we exponentiated them to find the rate parameter $\mathbf{y}_i$ of a Poisson process, which was sampled to obtain spiking data for $N$ neurons ($N=12$). These data were used only in Figure 2A.

\subsection{Real neural data \label{SuppBio}}

The maze dataset was previously collected from the motor cortex of a monkey performing a reaching task \cite{churchland_cortical_2010}. This dataset has been widely used to characterize the dynamics of motor cortical activity \cite{churchland_cortical_2010,pei_neural_2022, pandarinath_inferring_2018}. In particular, these data are well-modeled by autonomous dynamics \cite{pandarinath_inferring_2018}. 

The monkey was trained to perform a delayed reaching task in which it had to maintain its hand at the center of a 2D maze displayed on a screen while a target was shown somewhere within the maze. After a randomly-timed delay, a go-cue was issued which prompted the monkey to move its hand from the center of the screen to the indicated target. Each trial also had a set of obstacles (i.e., the walls of a maze) with various configurations that required the monkey to produce reaches with varied trajectories, even when they were directed towards the same target. A total of 108 of these maze configurations (i.e., target and obstacle combinations) are included in this dataset.

Neural activity was recorded using two Utah arrays \cite{maynard_utah_1997}, one in the dorsal premotor (PMd) cortex and one in the primary motor cortex (M1) \cite{churchland_cortical_2010}. Threshold crossings were sorted offline. The dataset contained 182 neurons in total, of which 137 were included in the held-in set and the remaining 45 were part of the held-out set. The held-out neurons were used to calculate the co-smoothing bits-per-spike metric (\ref{co-bps}). The monkey's hand and cursor positions were recorded during the experiment (\ref{vel}).

These data were downloaded from the Distributed Archives for Neurophysiology Data Integration (DANDI, \cite{rubel_neurodata_2022}). We binned spike counts at 20 ms and trialized and aligned the data to 250 ms before and 450 ms after movement onset. Further details can be found in \cite{churchland_cortical_2010, pei_neural_2022}.

\section{Model training}

\subsection{Simulated neural data}

All weights were initialized from \(\mathcal{U}(-\sqrt{k}, \sqrt{k})\), where \(k=1/\mathrm{in\_features}\) for linear layers and \(k=1/\mathrm{hidden\_size}\) for the GRU encoder weights. Dropout layers (\(p=0.05\)) were inserted before and after the initial condition linear projection during training. We used the average Poisson negative log-likelihood (NLL) across neurons and time points as our training objective. Models were trained incrementally to improve the stability of training: rather than compute loss on the whole trajectory, we added groups of 5 new time steps every 75 epochs, up to the max of 70 steps. Models were trained by stochastic gradient descent using Adam for 3000 epochs. A single learning rate was shared for the optimizer of the encoder, generator, and readout weights for each model. Each generator was a NODE that contained an MLP with six hidden layers, each with 128 ReLU units. 

\begin{figure}[ht!]
    \centering
    \includegraphics[width=5in]{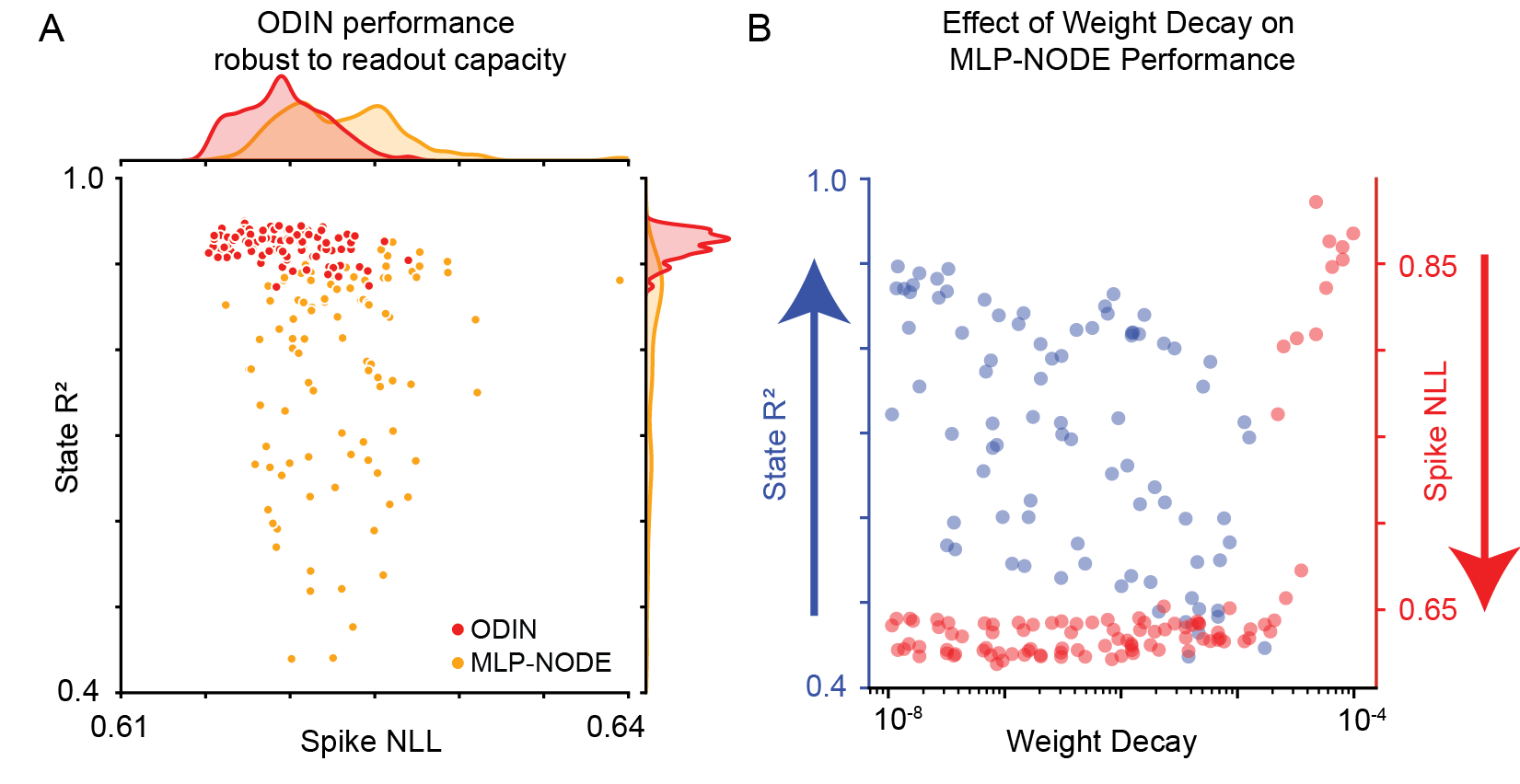}
    \caption{Example hyperparameter sweeps for ODIN and MLP-NODE}
    \label{fig:HPSweeps}
    \vspace{-0.25cm}
\end{figure}
 We performed initial hyperparameter sweeps to determine ranges that resulted in good reconstruction performance as measured by Spike NLL (see Methods), and used the same hyperparameter setting for models across state dimensionalities. Two example hyperparameter sweeps testing the effect of readout capacity (100 model initializations with readout hidden sizes in [60,200] and number of hidden layers in [1,3]) and weight decay (100 model initializations with weight decay drawn log-uniformly from [1e-8, 1e-4], Supp. Fig. \ref{fig:HPSweeps}). We found that across readout capacities, good reconstruction performance implied good latent recovery for ODIN but not MLP-NODE. Additionally, we found that increasing weight decay on MLP-NODE tended to degrade rather than improve latent recovery. Across all HPs tested, we found no hyperparameter settings for which ODIN had good reconstruction performance but poor latent recovery.

HPs for models trained on the Arneodo system are given in Table \ref{table:2}.

\begin{table}[h!]
\caption{Training hyperparameters (Synthetic Data)}
\centering
\begin{tabular}{|c|c c c|}
 \hline
 & & Arneodo &\\
 \hline
  & Linear & MLP & ODIN\\
 \hline
 Batch Size & 650 & 650 & 650 \\
 Learning Rate & 2e-3 & 1.88e-4 & 1.88e-4 \\
 Encoder Hidden Size & 100 & 100 & 100 \\
 Dropout & 0.05 & 0.05 & 0.05\\
 NODE Hidden Layers & 6 & 6 & 6 \\
 NODE Hidden Size & 128 & 128 & 128 \\
 Readout Hidden Layers & 0 & 2 & 2 \\
 Readout Hidden Size & - & 150 & 150 \\
\hline
\end{tabular}
\label{table:2}
\end{table}
\subsection{Real neural data}
The weight initialization procedure and dropout settings were the same as for the models trained on Arneodo. In addition to Poisson NLL, we also added regularization terms ($L_2$ norm on weights) and used different learning rates for the encoder, generator, and readout modules. We trained these models using Adam for 1500 epochs with the loss function given by Equation \ref{real_neural_data_loss}:
\begin{gather}
L(\mathbf{x}, \mathbf{\hat{y}}, \mathbf{\theta}_E, \mathbf{\theta}_G, \mathbf{\theta}_R) = \text{PoissonNLL}(\mathbf{x} | \mathbf{\hat{y}}) + \lambda_E ||\mathbf{\theta}_E||_2^2 + \lambda_G ||\mathbf{\theta}_G||_2^2 + \lambda_R ||\mathbf{\theta}_R||_2^2 \label{real_neural_data_loss}
\end{gather}
where $\mathbf{x}$ and $\mathbf{\hat{y}}$ represent the observed spiking activity and the predicted firing rates, respectively, and $\lambda _E, \lambda _G, \lambda _R$ represent the regularization coefficients for the $L_2$ regularization penalty applied to the model weights $\mathbf{\theta}_E, \mathbf{\theta} _G, \mathbf{\theta}_R$ of the encoder, generator, and readout, respectively. To improve training stability, we also used different learning rates for each component of the model ($\alpha _E, \alpha _G, \alpha _R$). Specific parameters for models trained on the Maze dataset are given in Table \ref{table:Maze}.

\subsubsection{AutoLFADS}
We trained AutoLFADS models of varying latent dimensionalities as a point of reference for ODIN's performance \cite{keshtkaran_large-scale_2022}. Notably, we used the autonomous version of LFADS and fixed the initial condition, generator, and factors dimensionality to $\hat{D}$ for these experiments. The batch size was 512 and the encoder hidden size was 100. Population-Based Training was used with a population of 20 workers to search initial learning rate (init: 1e-2, range: loguniform; 1e-5, 5e-2), dropout rate (init: 5e-2, range: uniform; 0.0, 0.6), coordinated dropout rate (init: 0.3, range: uniform; 0.01, 0.7), initial condition KL (range: loguniform; 1e-10, 1e-3), generator L2 scale (range: loguniform; 1e-10, 1e0), and encoder L2 scale (range: loguniform; 1e-10, 1e0). Linear ramp-up of KL and L2 penalties occurred over the first 80 epochs and the population was subjected to binary tournament and perturbation every 25 epochs for a total of 1000 training epochs.

\begin{table}[h!]
\caption{Training hyperparameters (Maze Data)}
\centering
\begin{tabular}{|c|c c|}
 \hline
 & \ \ \ \ \ \ \ \ \ \ \ \ \ \ \ \ \ \ Maze &\\
 \hline
  & Linear  & ODIN\\
 \hline
 Batch Size & 64 & 64  \\
 $\lambda _E$ & 1.6e-5 & 2.2e-6 \\
 $\lambda _G$ & 1.6e-5 & 1.35e-9 \\
 $\lambda _R$ & 1.6e-5 & 4.2e-6 \\
 $\alpha _E$ & 5e-3 & 4e-4 \\
 $\alpha _G$ & 5e-3 & 7e-4 \\
 $\alpha _R$ & 5e-3 & 1.4e-4 \\
 Encoder Hidden Size & 100 & 100 \\
 Dropout & 0.05 & 0.05\\
 NODE Hidden Layers & 6 & 6 \\
 NODE Hidden Size & 128 & 128 \\
 Readout Hidden Layers & 0 & 3 \\
 Readout Hidden Size & - & 128 \\
 Number of Flow Steps & - & 25 \\
 
\hline
\end{tabular}
\label{table:Maze}
\end{table}

\vspace{-3mm}
\section{Injectivity estimation}
\begin{figure}[ht!]
    \centering
    \includegraphics[width=3.58in]{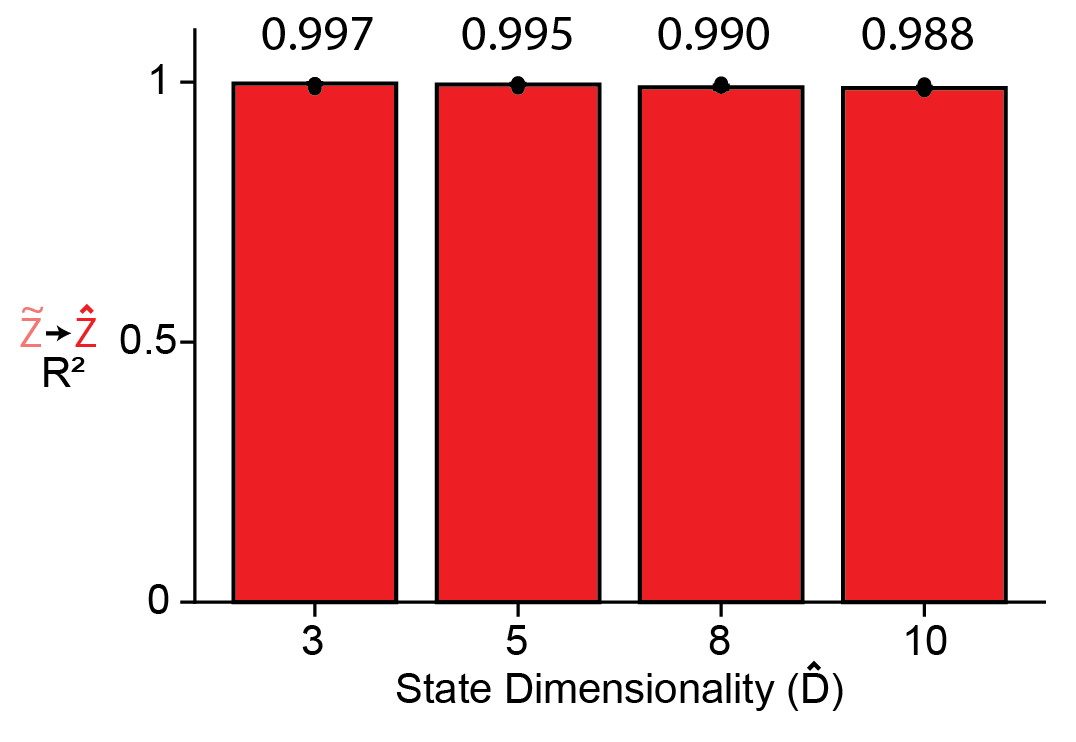}
    \caption{Injectivity of the Flow readout across state dimensionalities. Each bar indicates the mean value of 5 randomly initialized ODIN models for each state dimensionality. Results from individual models are plotted as points.}
    \label{fig:InjectionReverse}
    \vspace{-0.25cm}
\end{figure}

To demonstrate the approximate injectivity of the Flow readout, we tested whether the readout could be inverted to recover the inferred latent activity. The readout mapping $\hat{g}$ should satisfy the following equations
\begin{align}
    \Tilde{\mathbf{z}}_t &= \hat{g}^{-1}(\hat{g}(\hat{\mathbf{z}}_t)) \label{eq:tilde} \\
    \Tilde{\mathbf{z}}_t &\approx \hat{\mathbf{z}}_t
\end{align}
where $\hat{\mathbf{z}}_t$ is the inferred latent activity and $\Tilde{\mathbf{z}}_t$ is the latent activity recovered by the reverse pass of the Flow.

We computed the $R^2$ between the inferred and recovered $\mathbf{z}_t$ for these models and found that our mappings were able to recover the inferred $\hat{\mathbf{z}}_t$ with average $R^2$ values across randomly initialized models of 0.997, 0.996, 0.990, and 0.988 at $\hat{D} = 3,5,8,10$, respectively (Supplementary Figure \ref{fig:InjectionReverse}).

\subsection{Effective Rank}
To assess the injectivity of the Linear readout, we used a previously published method that determines the approximate number of significant singular values of a given matrix $A$ \cite{roy_effective_2007}. Let \(A\) be a complex-valued, non-all-zero matrix of size \(N \times \hat{D}\), where \(N > \hat{D}\) that acts as the weight matrix of a readout from inferred latents \(\hat{\mathbf{z}}\) to predicted log-rates \( \log \hat{\mathbf{y}}\) in the equation \(\log \hat{\mathbf{y}} = A \hat{\mathbf{z}} + \mathbf{b}\). We perform a singular value decomposition (SVD) on \(A\), such that \(A = U\Delta V\), where \(U\) and \(V\) are unitary matrices of size \(N \times N\) and \(\hat{D} \times \hat{D}\), respectively, and \(\Delta\) is an \(N \times \hat{D}\) rectangular diagonal matrix containing the real non-negative singular values \(\sigma_1 \geq \sigma_2 \geq \ldots \geq \sigma_{\hat{D}} \geq 0\).

For simplicity, let us define \(\sigma = (\sigma_1, \sigma_2, \ldots, \sigma_{\hat{D}})^T\). We then compute the singular value distribution \(p_k\), for \(k = 1, 2, \ldots, \hat{D}\), as

\begin{equation}
p_k = \frac{\sigma_k}{\|\sigma\|_1},
\end{equation}

where \(\|\sigma\|_1\) is the \(L_1\)-norm. Using this singular value distribution, we can calculate the Shannon entropy \(H\) as

\begin{equation}
H(p_1, p_2, \ldots, p_{\hat{D}}) = -\sum_{k=1}^{\hat{D}} p_k \log(p_k).
\end{equation}

The authors in \cite{roy_effective_2007} define the effective rank of the matrix \(A\), denoted as \(\text{{erank}}(A)\), using the Shannon entropy \(H\) as follows:

\begin{equation}
\text{{erank}}(A) = \exp\left(H(p_1, p_2, \ldots, p_{\hat{D}})\right).
\end{equation}

\begin{figure}[ht!]
    \centering
    \includegraphics[width=1.75in]{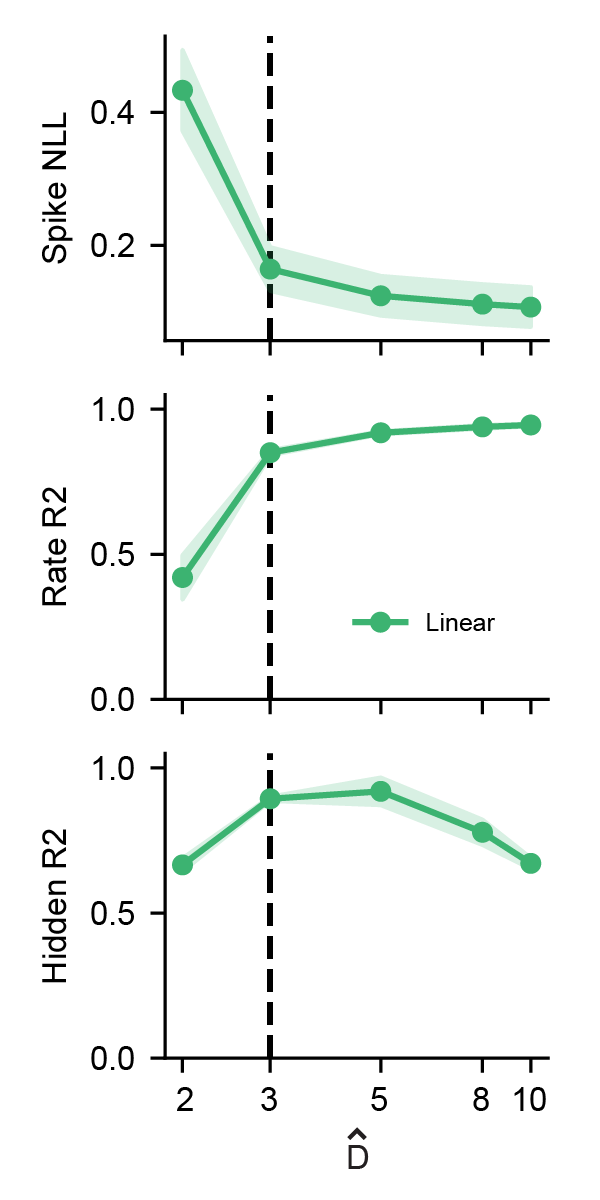}
    \caption{Linear-NODE trained on synthetic neural activity from linearly-embedded Arneodo system}
    \label{fig:NonWarpedArneodo}
    \vspace{-0.25cm}
\end{figure}

The effective rank gives us a measure of the number of significant singular values in \(A\). As traditional rank counts a matrix as being ``full-rank'' even if it has negligibly small but non-zero singular values, the effective rank provides a more informative assessment of the matrix's rank when used as the readout from a NPDM. We assessed the effective rank of the linear readout for 5 Linear-NODE models (with state dimensionality of $\hat{D}=2,3,5,8,10$, respectively) trained on synthetic neural data generated by linearly embedding trajectories from the Arneodo system (Section \ref{Arneodo-trajectories}) into log-firing rates, and found that while the reconstruction performance improved as $\hat{D}$ increased, the effective rank plateaued at erank $\approx 4$ (Fig 2A, Supp. Fig. \ref{fig:NonWarpedArneodo}).

\subsection{Cycle Consistency}
To directly compare injectivity of the Flow readout versus the MLP, we quantified how well each model's inferred latent activity could be recovered from the reconstructed log-rates. To do this, we took our fully-trained 10D ODIN and MLP-NODE models (shown in Fig. 2C, $\hat{D}=10$) and obtained the inferred latent activity $\hat{\mathbf{z}}$ and predicted log-firing rates $\log \hat{\mathbf{y}}$ from the Arneodo dataset. Then, we trained a separate network $h: \log \mathbf{Y} \rightarrow \mathbf{Z}$ to minimize the mean squared error between its output $\Tilde{\mathbf{z}}$ and the model-inferred latent activity $\hat{\mathbf{z}}$ (see Table \ref{table:CC-MLP} for hyperparameters). 
\begin{align}
\Tilde{\mathbf{z}} = h\left((\log \hat{\mathbf{y}})\right)
\end{align}
We computed the coefficient of determination between the re-generated latent activity $\Tilde{\mathbf{z}}$ and inferred latent activity $\hat{\mathbf{z}}$. If this performance is high, the inferred latents can be recovered from the log-rates suggesting that the readout is approximately injective.
\begin{align}
\text{Cycle Consistency} = R^2(\Tilde{\mathbf{z}}, \hat{\mathbf{z}})
\end{align}
\begin{table}[h!]
\caption{Training hyperparameters (Cycle-Consistency MLP, $h$)}
\centering
\begin{tabular}{|c|c|}
 \hline
 Parameter & Value \\
 \hline
 Batch Size & 2048 \\
 Learning Rate & 1e-3\\
 Hidden Layers & 3\\
 Hidden Size & 128\\
 Epochs & 1000\\
\hline
\end{tabular}
\label{table:CC-MLP}
\end{table}

It is possible for a readout to be fully injective (i.e., that $\hat{g}^{-1}$ exists), but still compress some features of latent activity into negligibly small contributions to the predicted firing rates, making the readout effectively, if not technically, non-injective. We reasoned that if this were the case, the inverse mapping $h$, in order to properly invert the warping applied by $\hat{g}$, would be highly sensitive to noise. We expect that such noise perturbations would be warped by $h$ into large changes in the predicted latents.  Using the models trained without noise, we computed the $R^2$ of re-generated latents $\Tilde{\mathbf{z}}$ compared to the inferred latents $\hat{\mathbf{z}}$. We therefore consider both the noise-free and noise-corrupted cycle consistency scores as indicators of the approximate injectivity of each readout, taking into consideration undue distortion applied in the process of learning the injective mapping.

\begin{equation}
\begin{gathered}
\Tilde{\mathbf{z}}_{\sigma} = h(\log \hat{\mathbf{y}} + \epsilon_{\sigma}), \quad \epsilon_{\sigma} = \mathcal{N}(0, \,\sigma), \ \ \sigma \in [0.01, 0.05, 0.1, 0.2, 0.5])\\
\end{gathered}
\end{equation}
\vspace{-3mm}
\begin{equation}
\begin{gathered}
ccR^{2}_{\sigma} = R^2(\Tilde{\mathbf{z}}_{\sigma}, \hat{\mathbf{z}})
\end{gathered}
\end{equation}

\vspace{-3mm}
\subsection{Alternative injective readout}
\vspace{-3mm}

\begin{figure}[ht!]
    \centering
    \includegraphics[width=1.75in]{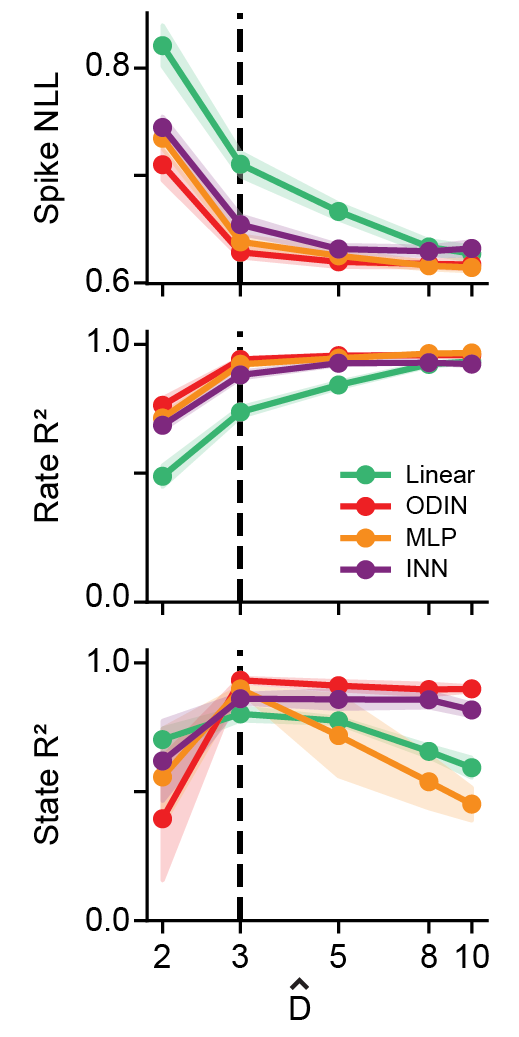}
    \caption{Invertible Neural Network readouts produce qualitatively similar results to Flow readout models. Data shown is the same as Fig. 2C, except overlaid with INN readout model (purple)}
    \label{fig:Invert}
    \vspace{-0.25cm}
\end{figure}

As an additional confirmation that injectivity was the critical addition to non-linear readouts that made latent recovery more robust, we tested an alternative injective architecture — an invertible neural network (INN) \cite{dinh2014nice}. We found that using a 6-layer INN in place of the Flow readout had comparable Rate $R^2$ and State $R^2$ to ODIN, and that, like ODIN, State $R^2$ was stable as $\hat{D}$ increased beyond $D=3$ (see Supp. Fig. \ref{fig:Invert}). Each INN layer was composed of coupling, permutation and affine transformations. Additional training parameters are noted in \ref{table:INN}. This result further supports our claims that injective networks empirically promote robust latent recovery. 

Unfortunately, the INN hidden layer size is obligated to be the size of either the input or output dimensionalities, whichever is larger. Therefore, in realistic biological datasets where the number of neurons can be highly variable across datasets, the capacity of the INN readout is intrinsically linked to the number of recorded neurons. For this reason, we chose to use the Flow readout, which decouples the computational capacity of the injective transformation from the dimensionality of the neural space.

\begin{table}[h!]
\caption{Training hyperparameters INN (Synthetic Arneodo Data)}
\centering
\begin{tabular}{|c|c |}
 \hline
  Parameter & Value\\
 \hline
 Batch Size & 650 \\
 Learning Rate & 1.88e-4 \\
 Encoder Hidden Size & 100 \\
 Dropout & 0.05\\
 NODE Hidden Layers &  6 \\
 NODE Hidden Size & 128 \\
 Readout Hidden Layers &  6 \\
 Readout Hidden Size & 12 \\
\hline
\end{tabular}
\label{table:INN}
\end{table}

\newpage
\section{Fixed point finding and characterization}

For each model (Linear-NODE, MLP-NODE and ODIN), we located fixed points (FPs) by finding the positions in the latent space that minimized the norm of the vector field via the objective $q = \frac{1}{2} \lVert \hat{f} \rVert_2^2$ \cite{sussillo_opening_2013, Golub2018}. We initialized our search with 1024 randomly sampled initial states from along inferred latent trajectories. We used Adam with a learning rate of 5e-2 to minimize the $q$-value for each point independently over 10,000 iterations. Candidate points that did not achieve a $q$-value less than a magnitude of 7e-3 were excluded. As more than one candidate can approach the same FP, we combined candidate points that were within a specified distance, $\epsilon$ = 1, from one another. In practice, points that were excluded had much larger $q$-values than the putative fixed points. We then linearized the dynamics around each FP and computed the system Jacobian to determine the stability and rotational character of the system around these FPs.

\section{Metrics}
\subsection{Synthetic data metrics}
\subsubsection{Rate reconstruction (Rate $R^2$) \label{rater2}}

We computed the coefficient of determination between true (\(\mathbf{Y}\)) and predicted (\(\hat{\mathbf{Y}}\)) rates for each neuron, and reported the average value across neurons.
\[
    \text{Rate }R^2 = R^2(\mathbf{Y}, \hat{\mathbf{Y}}) = \frac{1}{N}\sum_{i=0}^{N} 1 -  \frac{\sum(\mathbf{y}_i - \hat{\mathbf{y}}_i)^2}{\sum(\mathbf{y}_i - \bar{\mathbf{y}}_i)^2}
\]
\subsubsection{Latent state reconstruction (State $R^2$)} \label{stater2}

To compute State \(R^2\), we concatenated a vector of ones with the true latent states (\(\mathbf{Z_1}\)), then used the pseudoinverse to find the optimal affine transformation from the true latents to the inferred latents ($\hat{\mathbf{Z}}$) (i.e., optimal linear estimation). We computed the coefficient of determination (\(R^2\)) between the true and inferred latent activity with the same equation as in \ref{rater2}.
\begin{align}
    \mathbf{W}_z &= \mathbf{Z}_1^\dagger \hat{\mathbf{Z}} \\
    \text{State} \: R^2 &= R^2(\hat{\mathbf{Z}}, \mathbf{Z}_1 \mathbf{W}_z)
\end{align}
\subsubsection{Activation function comparison}
We developed a method for deriving an estimate of the inferred activation functions $\hat{\mathbf{\psi}}_i (\cdot)$ for a comparison to the true activation functions $\mathbf{\psi_i (\cdot)}$ (see Equation \ref{act_func}). We projected the true encoding vectors $\boldsymbol{\gamma} _i$ into the latent space of the model via the affine transformation $\mathbf{W}_z$ (see section \ref{stater2}). We then used these encoding vectors $\hat{\boldsymbol{\gamma}}_i \in \mathbb{R}^{\hat{D}}$ to convert inferred latent states $\mathbf{\hat{Z}} \in \mathbb{R}^{T \times \hat{D}}$ into an activation $\hat{\mathbf{a}}_i \in \mathbb{R}^{T}$ for each neuron.
\begin{align}
    \hat{\boldsymbol{\gamma}_i} &= \boldsymbol{\gamma}_{1,i} \mathbf{W}_z,  \ \text{for } i = 1, 2, \cdots, N\\
    \hat{\mathbf{a}}_{i} &= \hat{\mathbf{Z}} \cdot \hat{\boldsymbol{\gamma}_i}
\end{align}
To estimate the activation function for a given neuron $i$, we need pairs of inferred activations $\hat{\mathbf{a}}_{i}$ and firing rates $\hat{\mathbf{y}}_{i}$. For each neuron, we split firing rates into 20 quantiles and computed the corresponding median activation $\hat{\mathbf{a}}_{i,1:20}^{med}$ and firing rate $\hat{\mathbf{y}}_{i,1:20}^{med}$ within each quantile.
\begin{gather}
    \hat{\mathbf{y}}_{i,1:20}^{med}, \hat{\mathbf{a}}_{i,1:20}^{med} = \text{Quantize}(\mathbf{\hat{y}}_i, \mathbf{\hat{a}}_i, 20)
\end{gather}
We represented the inferred activation function $\mathbf{\hat{\psi}}_i (\cdot)$ using these activation-firing rate pairs. We then performed the same procedure on the true rates and activations to find a similar representation of the true activation function $\mathbf{\psi} _i (\cdot)$ for each neuron. To compare the true activation function $\psi (\cdot)$ to the estimated activation function $\hat{\psi} (\cdot)$, we combined the activations of each neuron $i$ and its corresponding firing rate as the columns of the matrices:
\[
\mathbf{\hat{\Psi}}_i = \begin{pmatrix}
\mathbf{\hat{a}}_{i}^{med} & \mathbf{\hat{y}}_{i}^{med} \\
\end{pmatrix}, \quad
\mathbf{\Psi}_i = \begin{pmatrix}
\mathbf{a}_{i}^{med} & \mathbf{y}_{i}^{med} \\

\end{pmatrix}
\]
Because the inferred latent activity can be scaled and translated arbitrarily with respect to the true latent activity, we found the optimal affine transformation between $\mathbf{\hat{\Psi}}_i$ and $\mathbf{\Psi}_i$. We used the $R^2$ of this mapping to quantify the correspondence between the two activation functions $\mathbf{\hat{\psi}}_i (\cdot)$ and $\mathbf{\psi}_i (\cdot)$ for each neuron.

\subsection{Neural Latents Benchmark metrics}
\subsubsection{Co-smoothing bits-per-spike (co-bps) \label{co-bps}}

A common failure mode of many dynamics models is to find latent activity that can accurately reconstruct the firing rates of neurons seen by the encoders, but fails to reconstruct neural activity of held-out neurons. To avoid this pitfall, we used a previously developed metric called co-smoothing bits-per-spike which evaluates reconstruction performance on a set of held-out neurons not visible to the encoders \cite{pei_neural_2022}. At a high-level, this metric quantifies how well the firing rates of the held-out neurons can be predicted from the spiking of the held-in neurons (see \ref{SuppBio}). This metric is defined by Equation \ref{co_bps_eqn} for each held-out neuron.
\begin{equation}
\text{co-bps} = \frac {1}{n_{s} \log 2} (\mathcal{L}(\mathbf{\hat{y}} _{n,t}; \mathbf{x}_{n,t}) - \mathcal{L}(\mathbf{\bar{y}}_{n,:}; \mathbf{x}_{n,t})) \label{co_bps_eqn}
\end{equation}
where $\mathbf{\bar{y}}_{n,:}$ is the mean firing rate for neuron $n$ across time, $n_s$ is the total number of spikes for that neuron, $\mathbf{\hat{y}} _{n,t}$ is the predicted firing rate from the model at time $t$, $\mathbf{x}_{n,t}$ represents the observed spiking of that neuron at time $t$, and $\mathcal{L}$ represents the Poisson log-likelihood. More information can be found in \cite{pei_neural_2022}.

\subsubsection{Velocity decoding $\mathbf{R}^2$ \label{vel}}
A common metric of performance is how well inferred firing rates can be used to predict behavioral variables, as this can be used downstream for decoding intent in clinical applications like brain-computer interfaces \cite{willett_high-performance_2021}. For the Maze dataset, hand velocity has been shown to be highly correlated with the neural firing in motor cortices. We compute this metric using the method from \cite{pei_neural_2022}, in which a ridge regression model is trained to predict the observed hand velocity from inferred firing rates. The coefficient of determination ($R^2$) was then evaluated on validation data that was not used to train the ridge regression velocity decoder. 

\section{Compute resources}

We used an internal computing cluster with a total of 30 Nvidia GeForce RTX 2080 Ti GPUs for model training. Each model trained on simulated neural data took approximately 3 hours to train, while each model trained on real biological data took approximately 1.5 hours to train. With 2 models training on each GPU, the 100 models included in Figs. 2, 3, 4, and 5 took approximately 150 GPU-hours and the 50 NODE-based models included in Fig. 6 took approximately 37.5 GPU-hours. FP finding was fast, requiring 1 minute for each model.
\section{Open-source packages used \label{packages}}

\begin{itemize}
  \item \href{https://github.com/pytorch/pytorch}{\texttt{torch}} \cite{paszke_pytorch_2019} (BSD license): Deep learning framework providing layer definitions, GPU acceleration, automatic differentiation, optimization, and more.
  \item \href{https://github.com/PyTorchLightning/pytorch-lightning}{\texttt{pytorch\_lightning}} (Apache 2.0 license): Lightweight wrappers for model training.
  \item \href{https://github.com/ray-project/ray}{\texttt{ray.tune}} \cite{liaw2018tune} (Apache 2.0 license): Distributed hyperparameter tuning.
  \item \href{https://github.com/williamgilpin/dysts}{\texttt{dysts}} \cite{gilpin_chaos_2021} (Apache 2.0 license): Implementations for modeled dynamical systems.
  \item \href{https://github.com/mattgolub/fixed-point-finder}{\texttt{fixed\_point\_finder}} \cite{Golub2018} (Apache 2.0 license): Inspiration for \texttt{torch}-based fixed point finder.
  \item \href{https://github.com/VLL-HD/FrEIA}{\texttt{FrEIA}} (MIT license): Implementation of alternative Invertible Neural Network architecture.
  \item \href{https://scikit-learn.org/stable/}{\texttt{scikit-learn}}\cite{scikit-learn} (BSD License): Implementations of linear regression models and principal component analysis. 
\end{itemize}

\newpage

\end{document}